\documentclass[EPJ,twocolumn]{webofc}
\usepackage[varg]{txfonts}   
\usepackage{subfigure}

\wocname{Atmohead Conference series}
\woctitle{Proceedings of the 2nd Atmohead Conference, May
  19-21, 2014. Padova (Italy) }
\begin{document}
\title{Thin and thick cloud top height retrieval algorithm with the Infrared Camera and LIDAR of the JEM-EUSO Space Mission}
%
%

\author{G. Sáez-Cano\inst{1}\fnsep\thanks{\email{lupe.saez@uah.es}} \and 
J. A. Morales de los Ríos\inst{1} \and 
L. del Peral \inst{1,2} \and
A. Neronov\inst{2}\and
S. Wada\inst{3} \and 
M. D. Rodríguez Frías\inst{1,2} \
for the JEM-EUSO Collaboration.
}

\institute{Space \& Astroparticle Group, UAH, Madrid, Spain
 \and
  ISDC, Astronomy Dept. University of Geneva, Switzerland.
 \and
  RIKEN Advanced Science Institute, Japan.
}

\abstract{%
The origin of cosmic rays have remained a mistery for more than a century. JEM-EUSO is a pioneer space-based telescope that will be located at the International
Space Station (ISS) and its aim is to detect Ultra High Energy Cosmic Rays (UHECR) and Extremely High Energy Cosmic Rays (EHECR) by observing the atmosphere.
Unlike ground-based telescopes, JEM-EUSO will observe from upwards, and therefore, for a properly UHECR reconstruction under cloudy conditions, a key element of
JEM-EUSO is an Atmospheric Monitoring System (AMS). This AMS consists of a space qualified bi-spectral Infrared Camera, that will provide the cloud coverage and
cloud top height in the JEM-EUSO Field of View (FoV) and a LIDAR, that will measure the atmospheric optical depth in the direction it has been shot. In this paper
we will explain the effects of clouds for the determination of the UHECR arrival direction. Moreover, since the cloud top height retrieval is crucial to analyze
the UHECR and EHECR events under cloudy conditions, the retrieval algorithm that fulfills the technical requierements of the Infrared Camera of JEM-EUSO to reconstruct 
the cloud top height is presently reported. 
}
\maketitle

\section{Introduction}
\label{intro}
Although cosmic rays have been discovered over one century ago, still their origin remains unknown. JEM-EUSO (Extreme Universe Space Observatory on-board the Japanese
Experiment Module) is a pioneer space-based experiment meant to detect Ultra High Energy Cosmic Rays (UHECR) and Extremely High Energy Cosmic Rays (EHECR). It will
observe the Extensive Air Showers (EAS) produced in the atmosphere when the UHECR and EHECR interact with the atmospheric nuclei.

To estimate the physical properties of the UHECR such as energy, arrival direction and its composition, the EAS fluorescence light profile and the EAS Cherenkov light
profile has to be measured. Since JEM-EUSO will cover a very large observation area in the atmosphere ($\sim10^5 km^2$), a properly monitoring of the atmospheric conditions
is mandatory. The Atmospheric Monitoring System (AMS) \cite{AMS} of JEM-EUSO will consist of a bi-spectral and space qualified Infrared Camera and a LIDAR
(LIght Detection And Ranging). The bi-spectral IR camera gives the cloud coverage in the FoV of the main telescope. With the difference in the brightness temperature
in the IR bands and a radiative retrieval model, the clouds altitude is provided. Thanks to the laser back-scatter signal, the optical depth ($\tau$) profiles of the
atmosphere in selected directions are measured (detecting aerosol and cloud layers).

To increase the statistics of the detected EAS, JEM-EUSO may use certain triggered events observed in cloudy scenarios \cite{saez1}. The main criteria to classify
cloudy events as "reconstructible events" is the detection of the depth of maximum development. The shower maximum will be observable
in cloudy conditions if $\tau < 1$ (the cloud is optically thin) or if $H_{max} > H_c$ (i.e., the altitude of the shower maximum is
higher than the cloud top height) \cite{saez2}, \cite{Adams}.

In this work we focus into two main features of the EAS reconstruction in cloudy conditions: the EAS geometry analysis and the
reconstruction of the cloud top height. 

\section{Shower geometry in cloudy conditions}
\label{sec}
EAS signal is detected as a spot moving on the focal surface of the JEM-EUSO main telescope. To analyse the direction of the incoming shower, the direction can
be understood as the composition of two projections: the azimutal angle, contained in the focal plane, and the zenith angle,
contained in the track detector plane (the plane which contains the shower track and the detector). The former is calculated
with the projection of the image of the shower in the focal surface. The latter, from the timing information and arrival angle
of the EAS photons to the detector \cite{abu}.

\begin{figure}[h!t] 
\begin{center}
\includegraphics[width=0.49\textwidth]{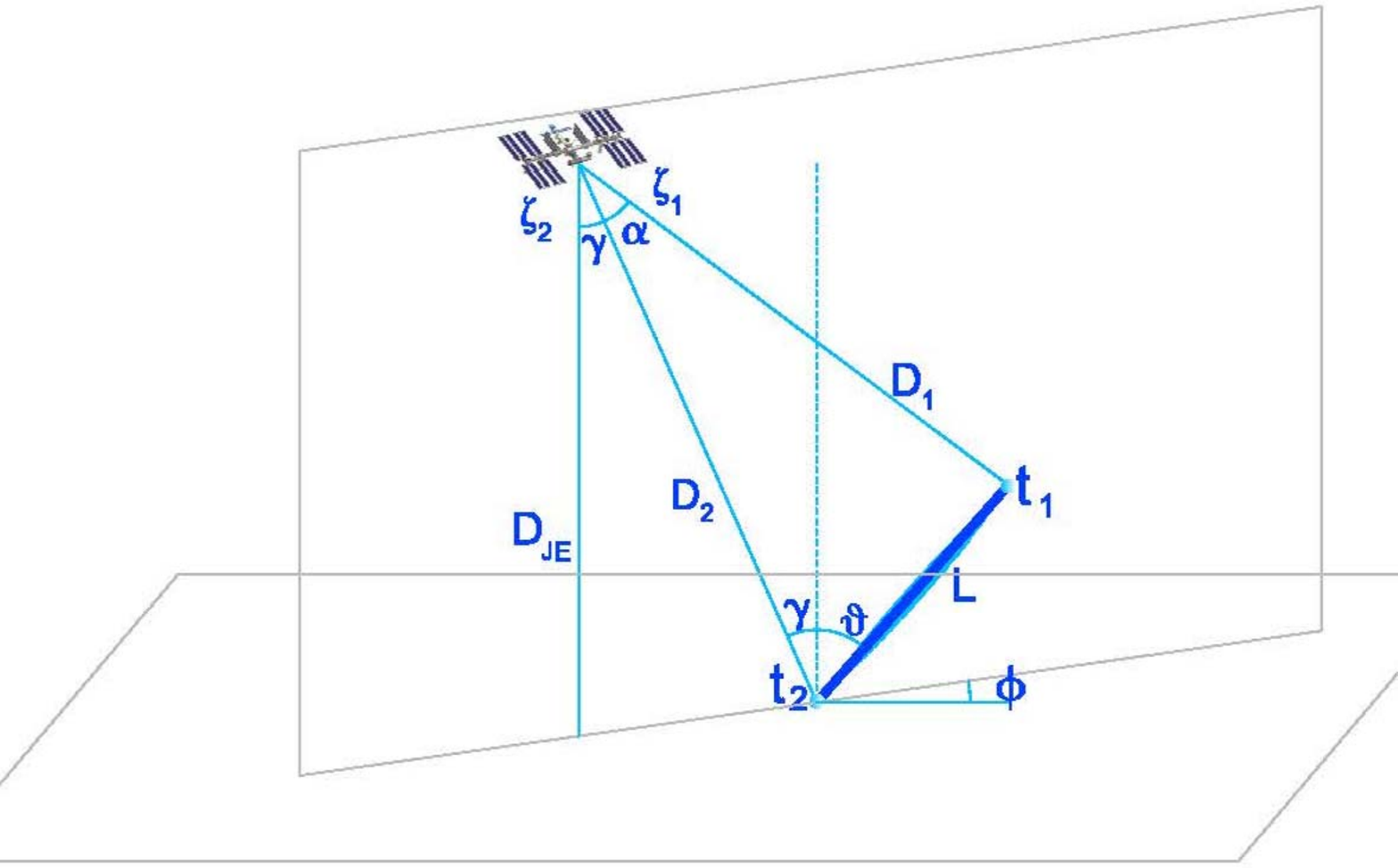}
\caption{\label{timing}Scheme of the geometrical relations used for the shower geometry.}
\end{center}
\end{figure}

The timing fit method is based on the viewing angle ($\alpha$), which is the angle between the direction given by two pixels, 
and the arrival time ($\zeta_{i}$). The zenith angle ($\theta$) can be calculated thanks to this timing fit (Figure \ref{timing}):\\

where $D_{JE}$ is the distance from JEM-EUSO to the ground, $D_{1}$ is the distance between the beginning of the shower to the 
telescope, $D_{2}$ is the distance from the telescope to the EAS core, L is the shower length, $\gamma$ is the angular distance
between the center of the FoV and the EAS core, and $t_{i}$ is the time at which the photons have been created. From the scheme
\ref{timing} it is clear that:
\begin{equation}
\label{eq1}
D_{2} = \frac{D_{JE}}{cos(\gamma)} \\ 
\end{equation}

\begin{figure}
\begin{center}
\subfigure{
\includegraphics[width=0.46\textwidth]{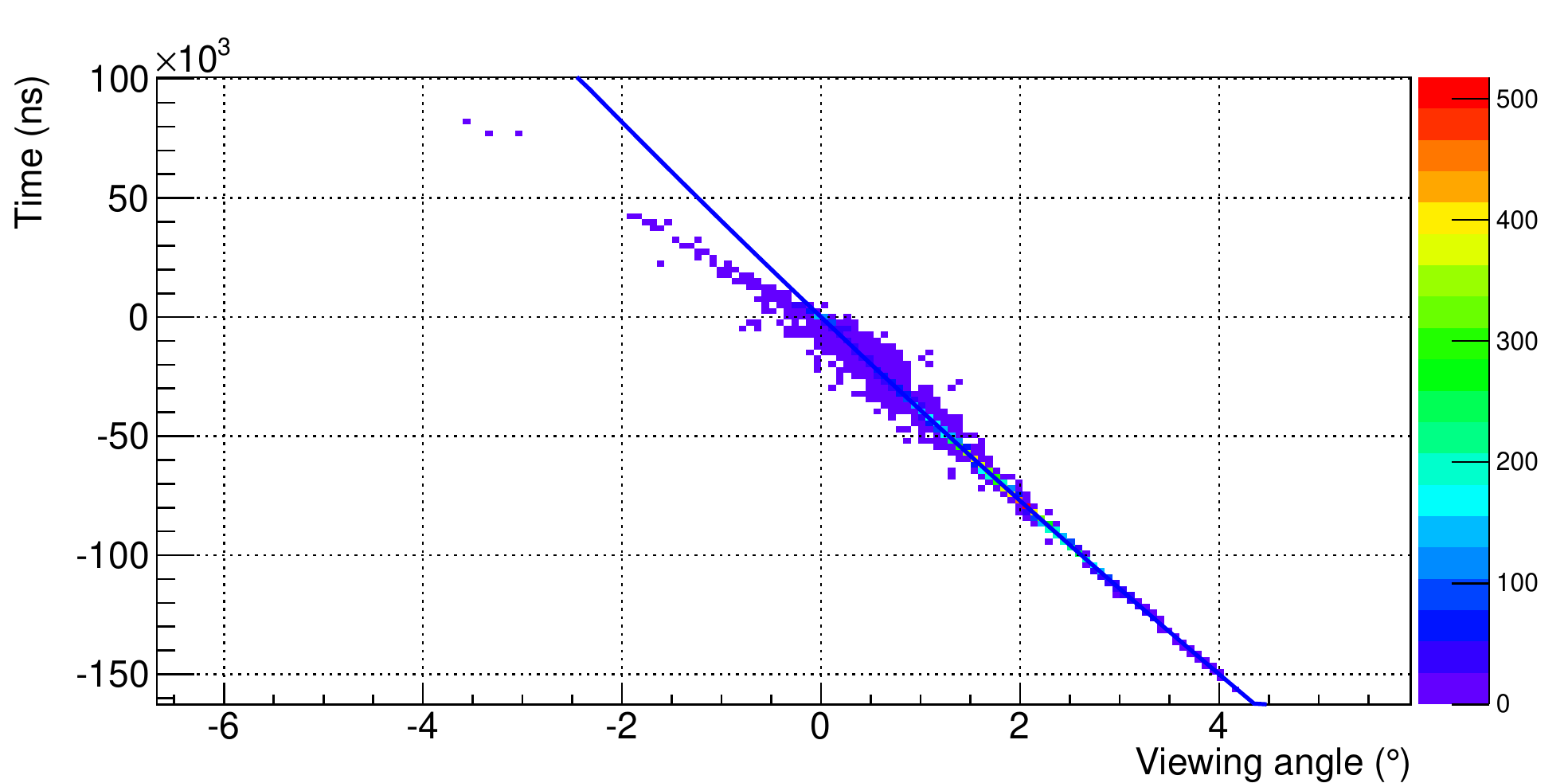}
}
\subfigure{
\includegraphics[width=0.46\textwidth]{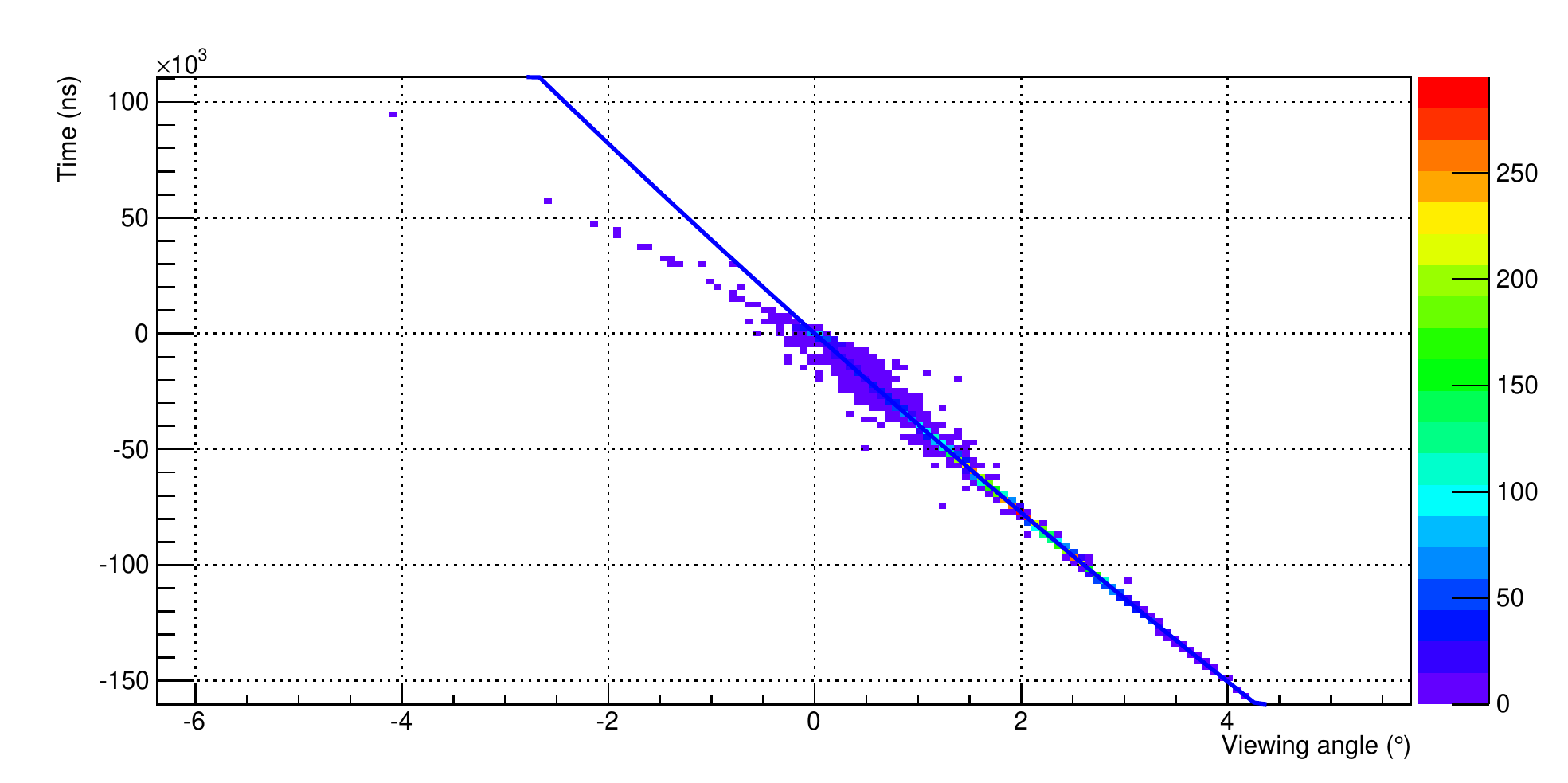}
}
\subfigure{
\includegraphics[width=0.46\textwidth]{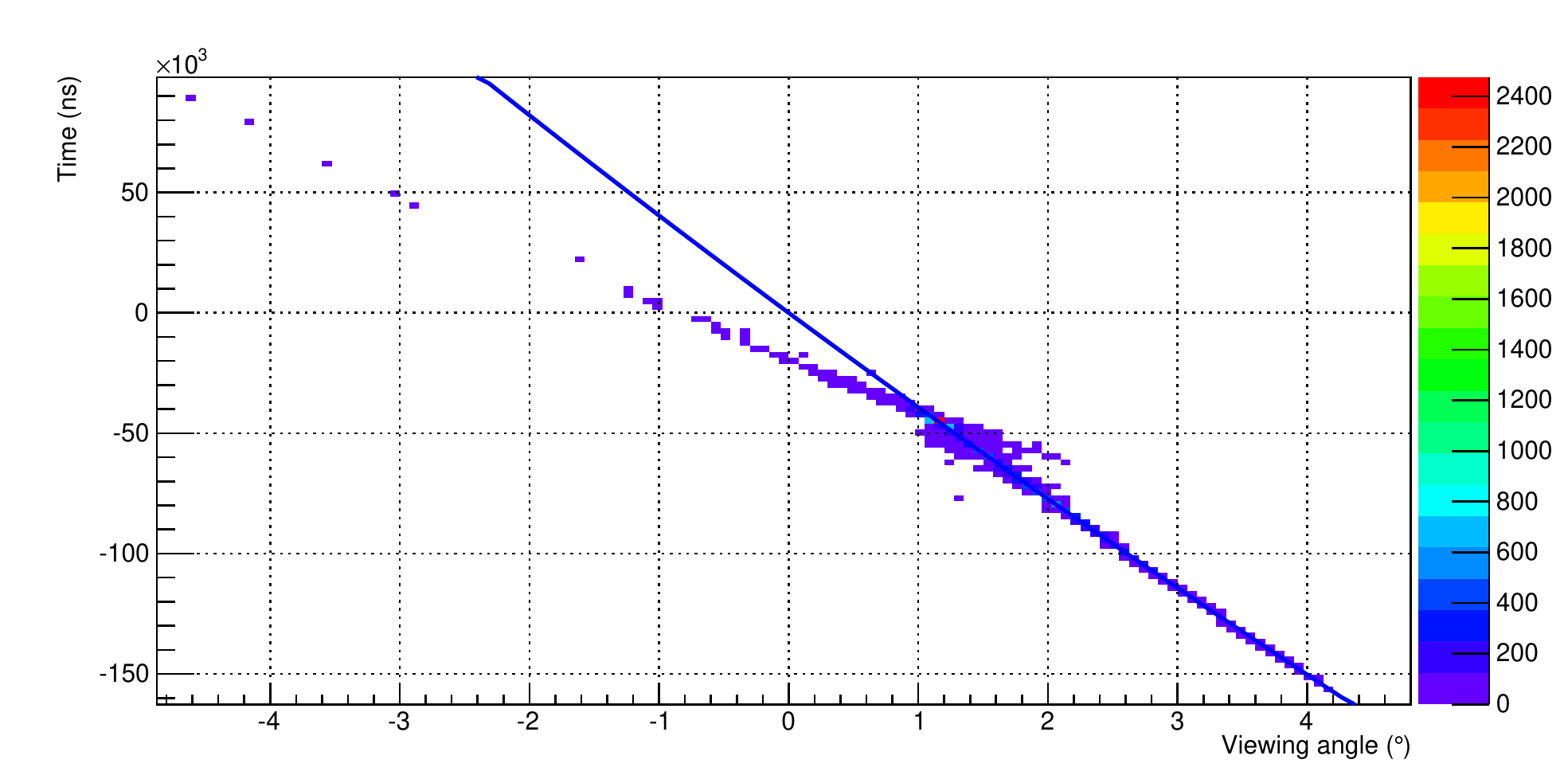}
}
\caption{\label{ct} Viewing angle as a function of time for three different cases: in clear atmosphere (upper panel), in presence
of an optically thin cloud (intermediate panel), and in presence of an optically thick cloud (lower panel). The slope of the fit is
the so called angular velocity. The color palette indicates the number of photons detected by each pixel.}
\end{center}
\end{figure}

\begin{figure}
\begin{center}
\subfigure{
\includegraphics[width=0.45\textwidth]{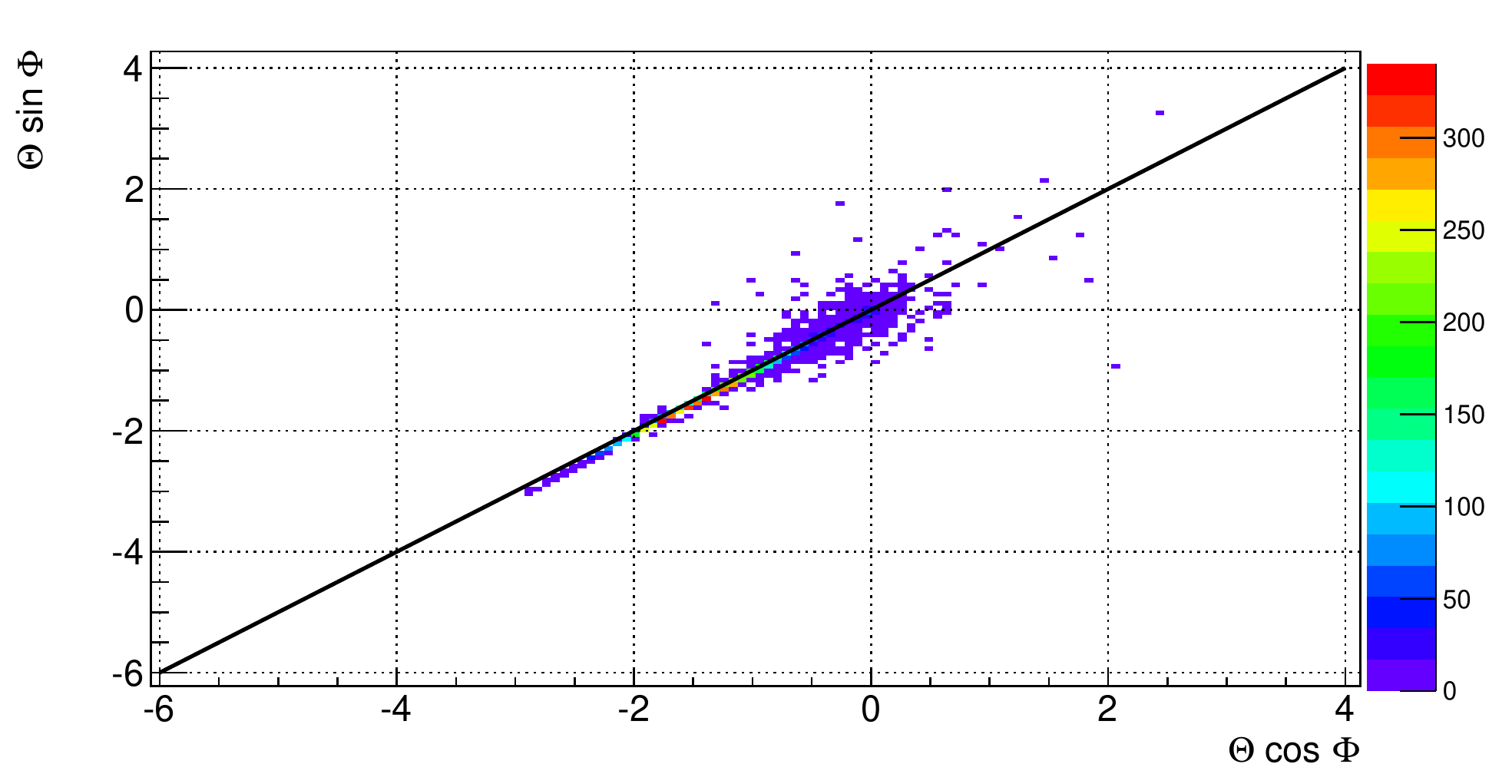}
}
\subfigure{
\includegraphics[width=0.45\textwidth]{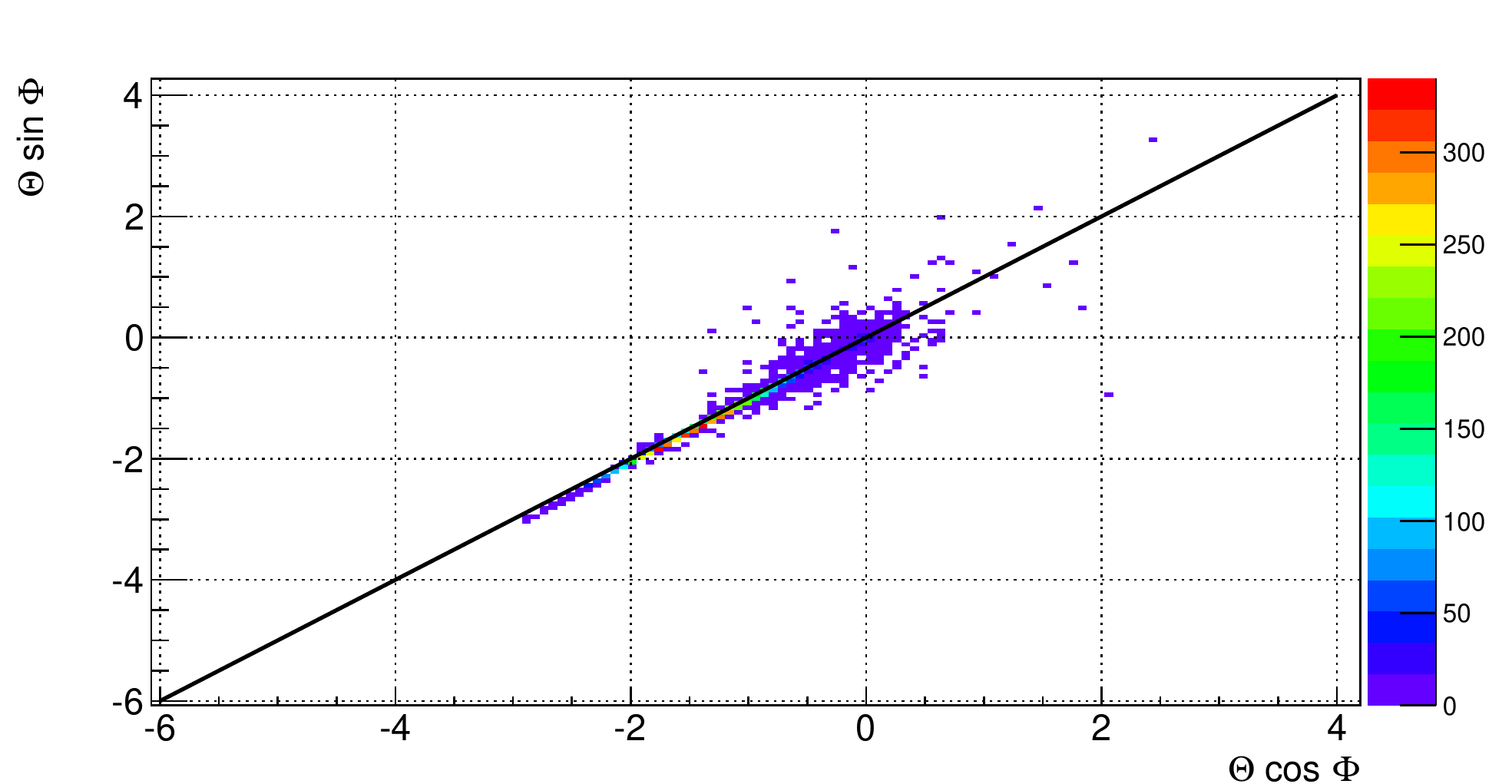}
}
\subfigure{
\includegraphics[width=0.45\textwidth]{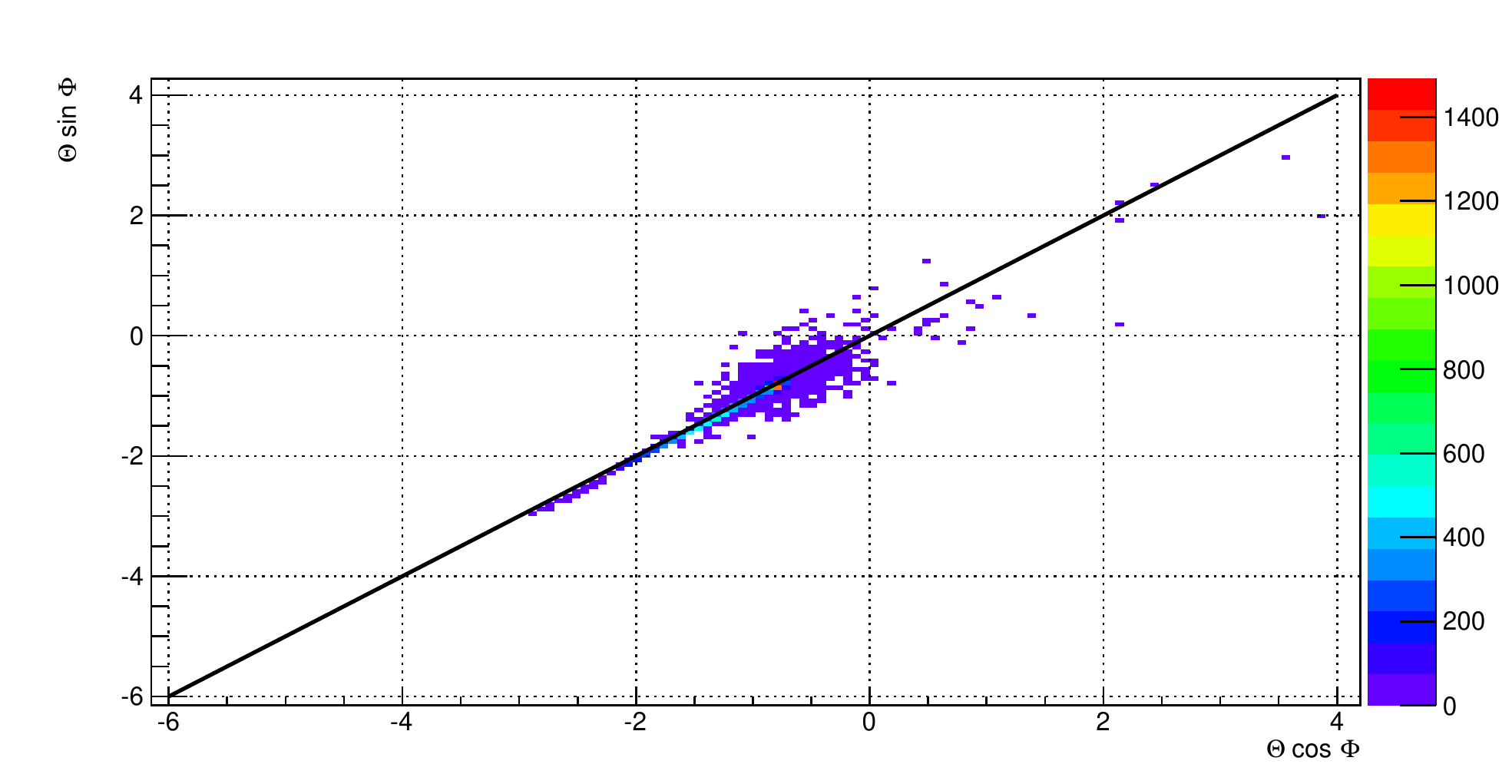}
}
\caption{\label{ct_phi} projection in the focal plane of the three previous cases: in clear atmosphere (upper panel), in presence
of an optically thin cloud (intermediate panel), and in presence of an optically thick cloud (lower panel). With the arctangent
this projection's slope, the azimutal angle is derived. The color palette indicates the number of photons detected by each pixel.}
\end{center}
\end{figure}

For simplicity we assume that the first EAS photon is emitted at $t_{1}=0$ seconds.
Then, the relation between the time and the zenith angle will be given by:

\begin{equation}
\label{tim}
 (\zeta_{1}-\zeta_{2})  =  \frac{D_{2}+l-D_{1}}{c}\\
\end{equation}

Which can be expressed as a function of $\gamma$, $\alpha$ and $\theta$:

\begin{eqnarray}
\label{geo3}
(\zeta_{1}-\zeta_{2}) & = & \frac{1}{c} \cdot \frac{D_{JE}}{cos(\gamma)} \cdot\nonumber\\
& & \left( 1 +\frac{sin(\alpha)-sin(\gamma+\theta)}{sin(\alpha+\gamma+\theta)}\right)\\\nonumber
\end{eqnarray}

For showers landing on the center of the FoV ($\gamma$=0), this equation is simplified to Eq. \ref{gammazero}.
\begin{equation}
\label{gammazero}
(\zeta_{1}-\zeta_{2}) = \frac{D_{JE}}{c} \cdot \left( 1 +\frac{sin(\alpha)-{sin(\theta)}}{sin(\alpha+\theta)}\right)\\ \\
\end{equation}

In Figure \ref{ct} it has been represented the viewing angle as a function of the arrival time for a proton induced shower with a
zenith angle of 60$^{\circ}$ and an azimutal angle of 45$^{\circ}$ landing on the center of the FoV in three different cases: in 
clear atmosphere (upper panel), in presence
of an optically thin cloud (intermediate panel), and in presence of an optically thick cloud whose cloud top height is slightly below the
shower maximum (lower panel). The signal has been fitted with the model \ref{gammazero}.
For the optically thin cloud case, even if some few photons have been scattered due to the cloud and their trajectory has changed,
the track of the shower is still well defined. On the other hand, in the case of the optically thick cloud the signal after the
cloud has been lost. However, a good fitting can be achieved with the part of the signal from above the cloud. 

In Figure \ref{ct_phi} the projection in the focal plane has been represented. To obtain the value of the azimutal angle, one needs
just to calculate the arctangent of the slope of this projection. As in the previous case, for the optically thin cloud case (intermediate
panel), even if some few photons have been scattered, shower track projection is still well defined. On the other hand, in the case
of the optically thick cloud (lower panel) the signal after the cloud has been vanished. However, a good fitting can be achieved 
with projection of the signal from above the cloud.

Thus, we can assume that angular reconstruction is feasible for the previously called "reconstructible" cases (i.e., the cloud is optically thin 
with $\tau < 1$, or the altitude of the shower maximum is higher than the cloud top height, $H_{max} > H_c$).

Nevertheless, in presence of optically thick clouds the angular resolution is modified. Since the signal after the cloud is totally lost,
the EAS shower track will be shortened, and therefore the angular resolution for an EAS in presence of an optically thick cloud will
be equivalent to that of an EAS in clear atmosphere with a lower zenith angle. On the other hand, the apparent movement of an EAS in
presence of an optically thin cloud will not be significantly modified, since the lenght of the shower track and its timing does
not change. Therefore, the angular resolution will be similar to that of the same EAS in presence of clear atmosphere.

\section{Cloud Top Height Retrieval algorithm}

We can parametrize the total extinction coefficient ($\epsilon$) in terms of the Ice Water Content (IWC) and on the fixed ice size distribution \cite{Liou}. 
Being $\beta_{ground}$ the irradiance of the ground at a certain temperature, and
$\beta_{cloud}$, the irradiance of the cloud at another temperature (following an atmospheric temperature profile), we can calculate
the detected irradiance by the IR-Camera as:

\begin{equation}
 \beta= \beta_{ground} \cdot (1-\epsilon) + \beta_{cloud} \cdot \epsilon
\end{equation}

\begin{figure}[h!t] 
\begin{center}
\includegraphics[width=0.51\textwidth]{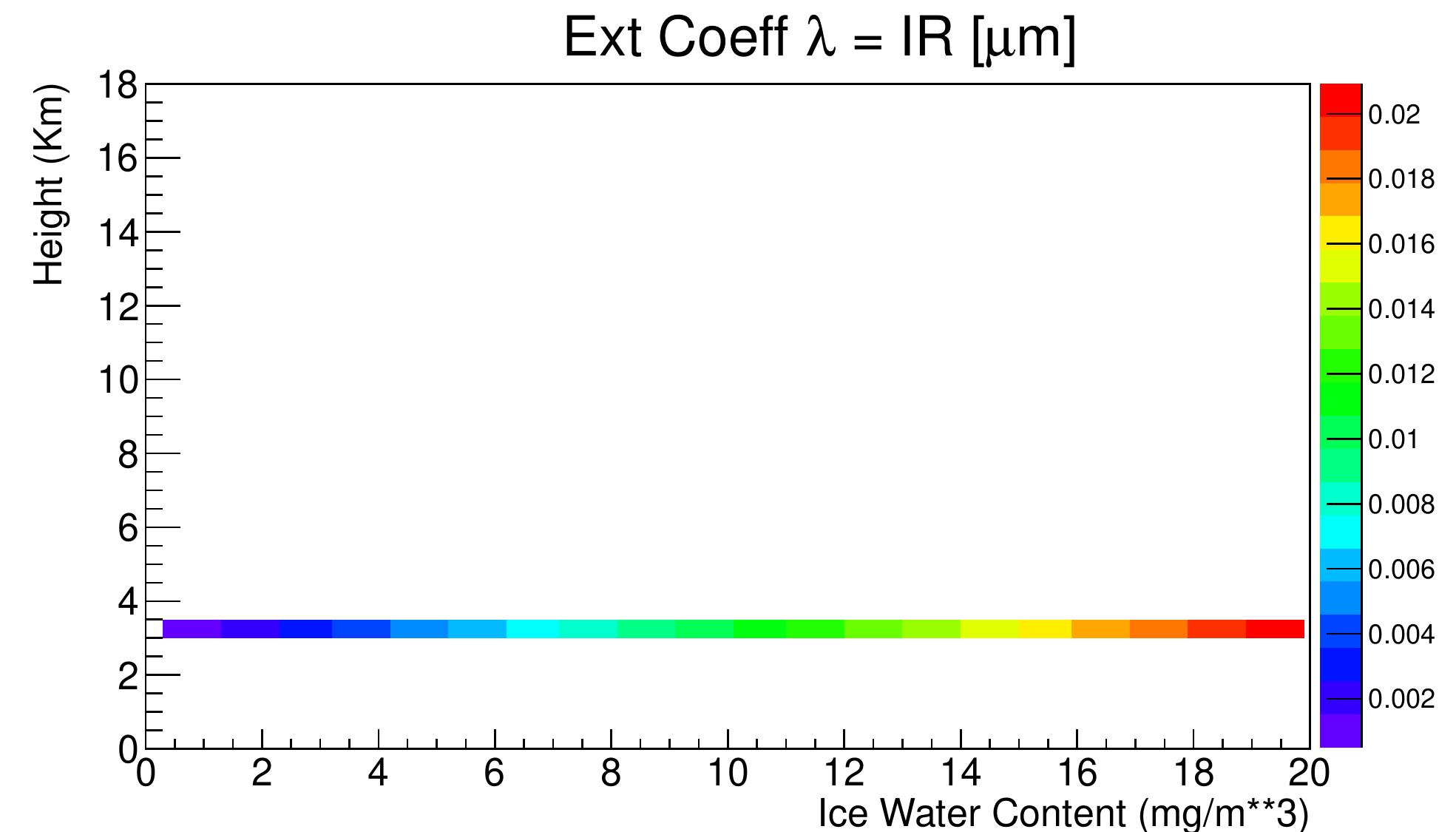}
\caption{\label{OD} Example of a cloud of 500 m thickness at 3 km altitude (Y axis). The figure represents the integrated extinction
coefficient in the near infrared (color scale) depending on the ice water content (X axis)}
\end{center}
\end{figure}

In Figure \ref{OD} it has been plotted, for a cloud of 500 m thickness at 3 km altitude, how the integrated extinction
coefficient in the near infrared varies with the ice water content of the cloud. Another cloud, has been modeled at 5 km, and the reconstruction performance is calculated
for both cases. In this paper, we have used an US-STD atmosphere temperature profile.

\subsection{LIDAR simulation}

\begin{figure}[h!] 
\begin{center}
\includegraphics[width=0.49\textwidth]{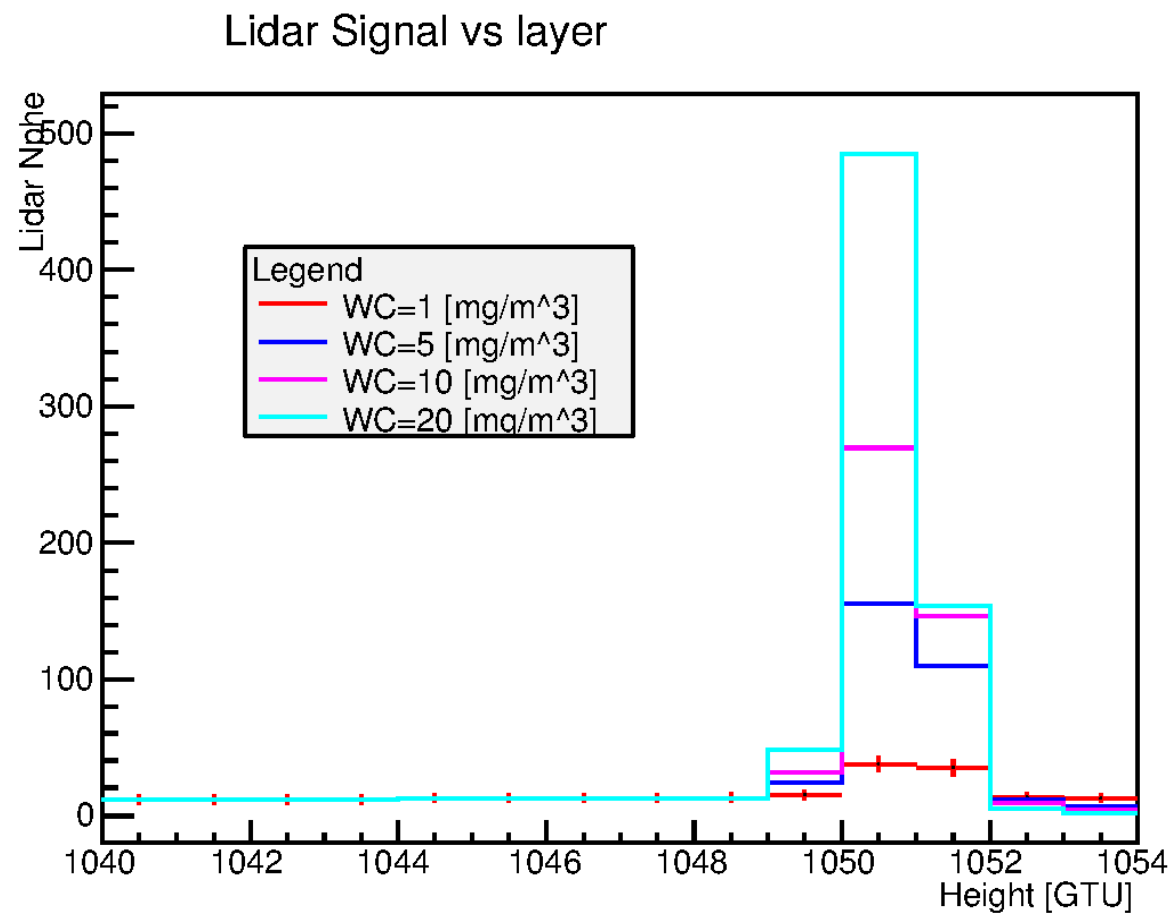}
\caption{\label{Lidar} Representation of the simulated backscatter LIDAR signal, at different IWC concentrations. Notice how the signal its spread for several GTU, this
leads that at different concentrations, and noise levels, the algorithm to detect the cloud from this signal has a resolution not lower than 2GTU.}
\end{center}
\end{figure}

For the same ice water content and the same fixed ice size distribution, we have simulated the extinction coefficient for the LIDAR wavelength,
and the backscattered signal that JEM-EUSO would detect to understand the capability of the LIDAR to measure the cloud optical depth \cite{lidar}.
Figure \ref{Lidar} shows an example of the simulated backscatter LIDAR signal, at different IWC concentrations for a cloud of
3 km altitude and a physical thickness of 500 m. In our simulation, the LIDAR resolution would be $\pm$ 2GTUs, which corresponds to
an altitude of 1500 m for a LIDAR signal that has been shot vertically. A more comprenhensive simulation and analysis of both LIDAR,
and Infrared Camera is under developement.

To model the atmosphere we have considered an hidrostatic approximation expressed in equation \ref{hid}:
\begin{equation}
\label{hid}
 \frac{d\rho}{dz}=-\frac{\rho}{g}
\end{equation}

Where $\rho$ is the density, $z$ is the height and $g$ is the gravity constant. If we integrate last equation, and replace $\rho$ with the ideal gas law definition $p= \rho \times R/M \times T$, the pressure can be written as:

\begin{equation}
 p=p_0 \times e^{\frac{M \cdot g \cdot z}{R \cdot T}}
\end{equation}

where $p$ is the pressure, $T$ is the temperature of the gas, $g$ is 9.80665 $m/s^2$, $M$ the air molar mass is 0.02897 kg/mol, $R$ the constant gas is 8.3144621 J/(mol$\times$k) and $p_0$ is the pressure at sea level that corresponds to 101325 Pa \cite{Liou}.

Therefore, we define the atmosphere pressure under the adiabatic aproximation as

\begin{equation}
 p \times T^{\frac{(\kappa-1)}{\kappa}}=Constant \\
\end{equation}

where $\kappa$ is the air heat capacity ratio, fixed as 7/5, considering the atmosphere a diatomic gas of only oxigen, and nitrogen. Combining the pressure equation in a gas enviroment, like the atmosphere, with the last equation of the adiabatic aproximation we obtain a temperture-height relation.

\begin{figure}[!htb] 
\begin{center}
\includegraphics[width=0.49\textwidth]{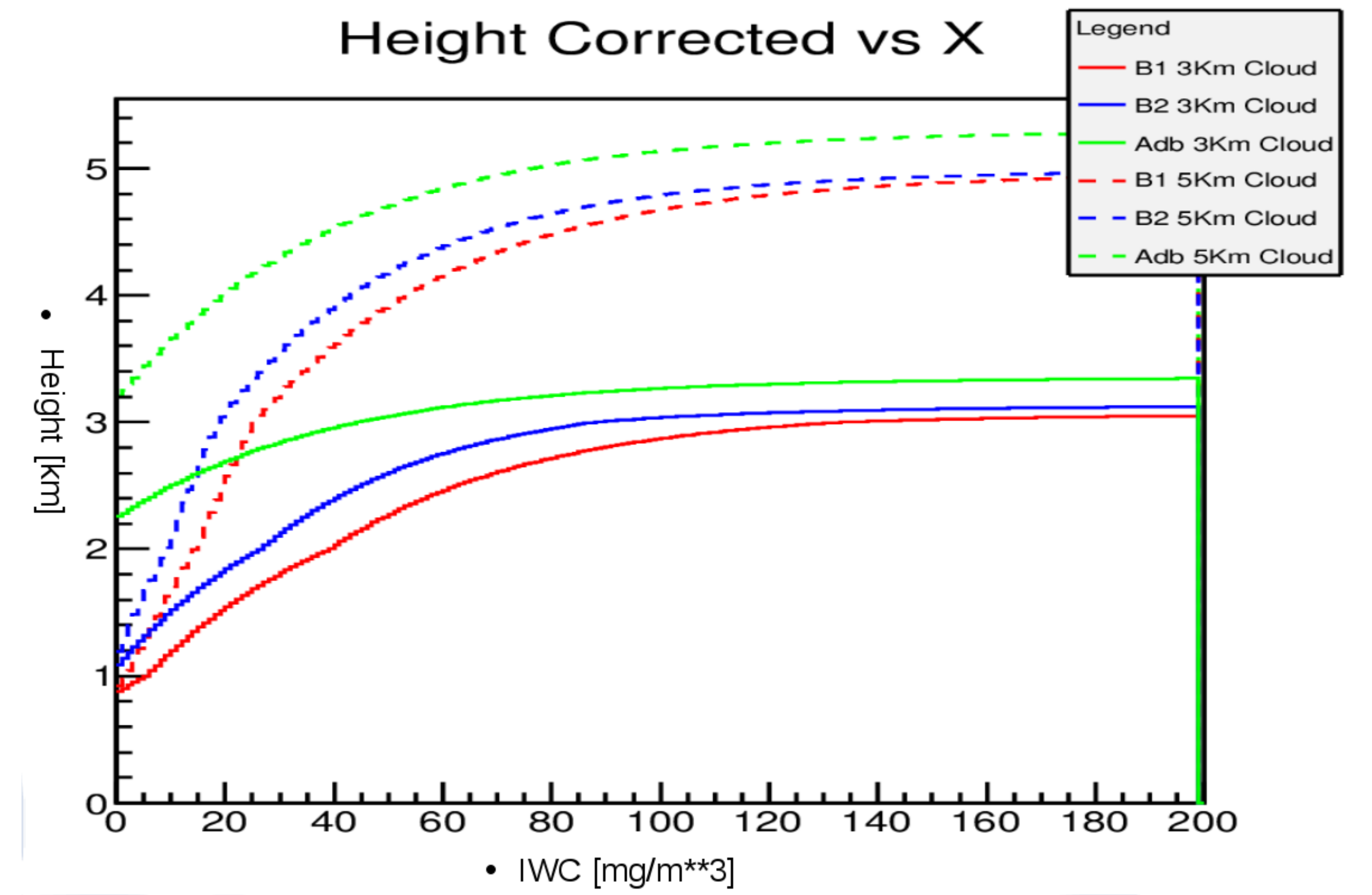}
\caption{\label{recoh} Retrieved height using the proposed abiabatic algorithm of a cloud at 3 Km and 5 Km. Comparison with the reconstructed height from the brightness
temperature in the B1=10.8 $\mu $ m, and B2=11.5 following the same atmospheric profile used in the simulation is presented as well.}
\end{center}
\end{figure}

As stated in the previous section, the LIDAR will provide us the altitude of the cloud area seen by a certain pixel, where the LIDAR has
been shot. Then we use this information, together with the temperature of the cloud retrieved by the Infrared Camera, to calculate the adiabatic constant that we need in our model.
To find out the altitude of the cloud seen by the rest of the IR camera pixels, we apply the adiabatic model with the obtained
adiabatic constant and we use the temperature of the different pixels. In the results presented in this paper, we have simulated a LIDAR shoot at an IWC=$100 mg/m^3$, then calculate the adiabatic constant and reconstruct the rest of the cloud.

\section{Summary and conclusions}

JEM-EUSO is an space based telescope which will use the atmosphere to detect the Extensive Air Showers (EAS) produced by the Ultra High Energy 
Cosmic Rays (UHECR) when they interact with atmospheric particles. Since UHECR have a very low flux, it is important to increase the exposure
of the telescope as much as possible. In this paper we prove how the arrival direction for events in presence of optically thin clouds
is similar to that of the same event occurred in a clear atmosphere. Also, it is explained why it is possible to obtain the arrival
direction of an EAS in presence of an optically thick cloud, if the shower maximum is visible, with a slightly worse resolution than the one
we would get for the same event in a clear atmosphere. 

To use events in cloudy conditions for our analysis, it is extremelly important to measure the properties of the clouds in the FoV of JEM-EUSO.
Moreover the radiometric retrieval algorithm that fulfills the technical requierements of the IR Camera of JEM-EUSO to obtain the top cloud height from the
temperature measured by the IR camera, using a single LIDAR shot for calibration is reported. However, as showed in this analysis, it is clear that the adiabatic
method aproximation is far more accurate, than the use of the brightness temperature and the atmosphere profile.  

\section{Acknowledgments}
The JEM-EUSO team at the University of Geneva acknowledges support from the Swiss Space Office through a dedicated PRODEX program.
This work is supported by the Spanish Government MICINN \& MINECO under projects projects AYA2009-06037-E/AYA, AYA-ESP 2010-19082, AYA-ESP 2011-29489-
C03, AYA-ESP 2012-39115-C03, AYA-ESP 2013-47816-C4, MINECO/FEDER-UNAH13-4E-2741, CSD2009-00064 (Consolider MULTIDARK) and by Comunidad de
Madrid under projects S2009/ESP-1496 \& S2013/ICE-2822. The calculations were performed using the RIKEN Integrated Cluster of Clusters
(RICC) facility \& the Space and Astroparticle SPAS-UAH Cluster. J.A. Morales de los Rios wants to acknowledge the financial support from the UAH-FPI grant
and the RIKEN-IPA program. M. D. Rodriguez Frias acknowledge International Visitor Grant from the Swiss National Science Foundation (SNSF).

\end{document}